\documentstyle[prl,twocolumn,aps,epsf]{revtex}
\begin{document}
\renewcommand{\epsffile}{1}{Missing author supplied figure}

\title{
QUANTUM COHERENCE IN MAGNETIC GRAINS }

\author{N. V. Prokof'ev$^{1,2}$  and P. C. E. Stamp$^{2}$}
\address{
$^{1}$ Russian Research Center "Kurchatov Institute", Moscow 123182, Russia\\
$\;\;\;$ \\
$^{2}$ Physics Department, University of British Columbia, 6224 Agricultural
Rd.,\\
 Vancouver B.C., Canada V6T 1Z1 \\ }
\maketitle

\begin{abstract}
We consider a model of the dynamics of a magnetic grain, incorporating
interactions with nuclear and paramagnetic spins, conduction electrons, and
phonons. Decoherence comes both from the spins and the electrons,
although electron effects can be neglected if the substrate is insulating, and
the grain has spin $S \le 10^5$. A successful experiment will require
both isotopic and chemical purification, if coherence is to be seen;
we suggest a "spin echo" arrangement to do this.
\end{abstract}

\bigskip

Much activity in the new field of nanomagnetism \cite{1} focuses on the
quantum dynamics of small magnetic particles or wires \cite{2,3}.
Of potential interest to the magnetic recording, information storage, and
computer industries, this work also seeks to understand quantum magnetism
on the mesoscopic scale - practical applications are still unclear. Recently
a debate has arisen over experiments \cite{4,5} claiming to see quantum
coherent motion of the magnetisation in ferritin nanomolecules. Criticisms
have centered around, e.g., (i) the dipolar coupling between molecules
\cite{2}, (ii) the experimental power absorption \cite{6,7}, (iii) decoherence
or
coherence blocking by environmental spins \cite{8,9,10}, and (iv) the
apparent inconsistency with high-T magnetic blocking \cite{11}. Nevertheless
recent experiments \cite{5} see further evidence of coherence.

Here we present a theory of coherence in mesoscopic magnetic molecules or
grains, including environmental spins, phonons, and possible electrons, and
also
the coupling to the measuring system. This has  implications for the ferritin
experiments; more importantly, it  shows what has to be done in future
experiments  to see coherence.

Many of our results can be understood qualitatively by simple arguments. At low
$T$, one models the grain as a "giant spin" $\vec{S}$, with effective
Hamiltonian truncated to the 2-level $\hat{H}^o_{eff} =2\Delta_o \hat{\tau}_x
\cos \pi S$, involving coherent motion between, e.g., the 2 quasiclassical
states $\vec{S}=\pm \hat{\vec{z}} S$ at a frequency $4 \Delta_o$ when
symmetry permits \cite{2,12}. To maintain
coherence  one requires (i) near degeneracy, within energy $\sim \Delta_o$, of
the 2 states, and (ii) no phase decoherence.

Consider now a Hamiltonian
\begin{equation}
H_{Hyp}={1 \over S}\bigg[ (-K_{\parallel}\:S_z^2 + K_{\perp}S_y^2)+
\sum_{k=1}^N \omega_k \vec{S} \cdot {\vec I}_k \bigg] \;,
\label{1}
\end{equation}
with hyperfine couplings $\{ \omega_k \}$ between $\vec{S}$
and nuclear spins
$\{ \vec{I}_k \}$; with no nuclear spins $\Delta_o = \Omega_o e^{-A_o}$,
where $\Omega_o \sim 2(K_{\parallel} K_{\perp})^{1/2}$, and $A_o \sim S
(K_{\parallel} /K_{\perp})^{1/2}$. Typically $\Omega_o \sim
10^{9}-10^{11}\:Hz$,
and $\omega_k \sim 10^{7}-10^{9}\:Hz$; for mesoscopic spins ($S \ge 10^3$), one
expects $\Delta_o < 10^{6}\:Hz$, for tunneling between $\pm \hat{\vec{z}}S$,
so $\Delta_o \ll \omega_k$.

However, even if the applied field $\vec{H}_o=0$, an internal bias field
$\epsilon = \sum_{k=1}^N \omega_k {\vec I}_k$ acts on $\vec{S}$. Suppose we
have
a dominant hyperfine coupling $\omega_o$, with a spread $\delta \omega_k$ of
levels around $\omega_o$, caused by Nakamara-Suhl interactions
($\delta \omega_k \sim 10^{3}-10^{6}\:Hz$), transfer hyperfine couplings
($\delta \omega_k \sim 1-20\:MHz$) or other magnetic hyperfine couplings.
Define a "density of states" $W(\epsilon )$ for the bias; if
$\mu = N^{1/2}\delta \omega_k/ \omega_o \ll 1$, this consists of
"polarisation groups" of
width $\sim \mu \omega_o $ around $\epsilon =
\omega_o \Delta N $, where $\Delta N =
N^{\uparrow} - N^{\downarrow}$ is the total nuclear polarization \cite{13},
inside a Gaussian envelope of width $N^{1/2} \omega_o$. Usually however
$\mu \gg 1$, and different polarization groups overlap.
Notice only the fraction $A$ of grains in an ensemble having
$\epsilon \le \Delta_o$ can flip - the rest are "degeneracy blocked". Thus
if $\mu \ll 1$, a fraction $\Delta_o/(N^{1/2}\delta \omega_k)$ of that
portion $f = \sqrt{2/ \pi N} $ of grains having $\Delta N=0$ can flip, so
$A \sim  \Delta_o/(N\delta \omega_k)$;
if $\mu \gg 1$, then  $A \sim \Delta_o/(N^{1/2}\omega_o) $.

Even these few grains may not necessarily flip {\it coherently}.
Flipping
$\vec{S}$ causes transitions in the nuclear system, which destroy phase
coherence, even if the change in polarisation
$\Delta M =0$; and other fields can cause a mismatch
between initial and final state nuclear wave-functions, analogous to the
Anderson orthogonality catastrophe \cite{14}. Electrons or
phonons can remove the degeneracy blocking, by absorbing the bias energy,
but further destroy coherence in doing so. Finally, nuclear spin diffusion, at
a rate $T_2^{-1}$, destroys coherence by causing
$\epsilon$ to fluctuate in time.

To see coherence is clearly going to be hard; we now proceed to a
quantitative theory, to see what can be done.

(i) \underline{The Spin Environment}: At energies $\ll \Omega_o$, spin effects
are treated by truncating to an effective Hamiltonian $H_{eff} = H_{SN}
+H_{NN}$, where \cite{8,9,10}
\begin{eqnarray}
&  H_{SN}&= 2{\tilde \Delta}_o \big\{ {\hat \tau }_+\cos \big[ \Phi   +
\sum_{k=1}^N
( \alpha_k {\vec n}_k -i \xi_k {\vec v}_k ) \cdot {\hat {\vec \sigma }}_k
\big] +H.c. \big\} \nonumber \\ &+&
{\hat \tau }_z \sum_{k=1}^N {\omega_k^{\parallel} \over 2}  \:
 {{\vec l}_k \cdot {\hat {\vec \sigma }}_k } +
\sum_{k=1}^N {\omega_k^{\perp} \over 2}  \:
 {{\vec m}_k \cdot {\hat {\vec \sigma }}_k } \;,
\label{2}
\end{eqnarray}
\begin{equation}
H_{NN} =  {1 \over 2} \sum_{k\ne k^\prime } V_{k k^\prime }
^{\alpha \beta }
 \hat{\sigma}_k^\alpha   \hat{\sigma}_{k^\prime }^\beta  \;.
\label{2n}
\end{equation}
$H_{SN}$ describes all couplings between $\vec{S}$ (now $\vec{\tau}$) and
the spin-$1/2$ environmental variables $\{ \vec{\sigma}_k \}$. The
dimensionless
parameters  $\alpha_k$,  $\xi_k$, and $\Phi$, the energies
${\tilde \Delta}_o $, $
\omega_k^{\parallel}$, and $\omega_k^{\perp}$, and the unit vectors
$\vec{n}_k$, $\vec{v}_k$, $\vec{l}_k$, $\vec{m}_k$, have been derived for
various "bare Hamiltonians". If, e.g., the $\{ \vec{I}_k \}$ in (\ref{1}) are
spin-$1/2$ nuclei,  one finds ${\tilde \Delta}_o =\! \Delta_o$,
$\Phi = \! \pi S$, $\xi_k=\alpha_k/\sqrt{2} =
\pi \omega_k /2 \Omega_o $ for small $\omega_k /\Omega_o$ (for general
$\omega_k /\Omega_o$, see Refs.\cite{9,10}), $\omega_k^{\parallel}=\omega_k$,
and $\omega_k^{\perp}=0$, whilst $\vec{n}_k=2^{-1/2} (\hat{\vec{x}},
\hat{\vec{y}})$, $\vec{v}_k=-\hat{\vec{x}}$, $\vec{l}_k=\hat{\vec{z}}$, and
$\vec{m}_k$ is arbitrary.
The "longitudinal coupling"
$\omega_k^{\parallel}$ gives the change in energy of $\vec{\sigma}_k$ when
$\vec{S}$ flips, and $\omega_k^{\perp}{\vec m}_k$ represents any field on
$\vec{\sigma}_k$, perpendicular to $\vec{l}_k$, which is unchanged by
tunneling;
this arises if $\vec{\sigma}_k$ comes from truncating a higher-spin
system (e.g., if $I_k$ is spin $1$), or if
the 2 quasi-classical orientations
of $\vec{S}$ are not exactly antiparallel. In reality $\alpha_k \ll 1$, and is
just the amplitude for $\vec{\sigma}_k$ to coflip with $\vec{S}$; when
$\vec{S}$ flips, roughly $\lambda = 1/2 \sum_k \alpha_k^2$ nuclei also flip.

To study coherence we calculate  $P(t) = \langle \hat{\tau}_z(t)
\hat{\tau}_z(0) \rangle $ \cite{15}, which shows coherent oscillations without
the environment (i.e.,
$P(t) \to P^{(0)}(t) = \cos [4\Delta_o t \cos \pi S ]$). We have solved for
$P(t)$, including all terms in (\ref{2}) and (\ref{2n}). We write it as
\begin{equation}
 P(t) =1- \sum_{M} \int d\epsilon {W(\epsilon )e^{-\epsilon /T}
\over Z(T)}
P_M(t,\epsilon -M\omega_o/2 )\;,
\label{6}
\end{equation}
with $W(\epsilon )$ as before, and  $Z(T)$ the nuclear partition function.
Eq.(\ref{6}) averages over bias $\epsilon $ and sums over processes in
which the polarisation changes by $2M$. The case $\omega_k^{\perp} = \xi_k
 =0$ is sufficient \cite{16} for what follows; then
\begin{eqnarray}
 P_M(t) =&& \int_0^\infty dye^{-y} \sum_{\nu=-\infty}^{\infty} \int {d\varphi
\over 2 \pi } F_{\lambda^\prime }(\nu ) \nonumber \\
&\times & e^{2i\nu (\Phi -\varphi )} P_M^{(0)} (t,\epsilon ,\varphi ,y)  \;,
\label{9}
\end{eqnarray}
where if $T_2^{-1} =0$ (no spin diffusion), we have
\begin{equation}
 P_M^{(0)} =  {\Delta_M^2(\varphi ,y) \over E_M^2 (\varphi ,y) }
\sin ^2 (E_M^2 (\varphi ,y)t) \;,
\label{6new}
\end{equation}
describing a biased 2-level system, where $E_M^2 = \epsilon^2 + \Delta_M^2$,
and $\Delta_M(\varphi ,y)=  2{\tilde \Delta}_o \mid \cos \varphi J_M
(2\sqrt{(\lambda -\lambda^\prime)  y}) \mid $, with $\lambda^\prime  = 1/2
\sum_k \alpha_k^2 (n_k^z)^2$ (so that $\lambda \ge \lambda^\prime$).
The averages over topological phase $\varphi$,
winding number $\nu$ (cf. \cite{8}), and "orthogonality blocking"
$y$, then describe all the dynamical effects of the $\{ \vec{\sigma}_k \}$ on
$\vec{S}$. The weighting factor $F_{\lambda^\prime }(\nu )= \exp \{
-4\lambda^\prime \nu^2 \}$; often, as for (\ref{1}), $n_k^z =0$, so
$ \lambda^\prime =0$ and the phase average collapses to a delta-function
$\delta (\Phi -\varphi )$.

In rare cases $\lambda \ll 1$ (i.e., no nuclei flip),
and (still assuming $T_2^{-1} =0$) the nuclear bath acts as a static field.
An example may be ferritin, where only $2.16$\% of the $Fe$ nuclei have
spin-$1/2$ (so $N\sim 100$), and  $\omega_o \sim 50\: MHz$
(but $\mu >1$ if we take account of hyperfine interaction with
$\sim 5000$ H$^1$ nuclei). Assuming \cite{4} that
$\Omega_o \sim 4\times 10^{10} Hz$, one gets
$\lambda \sim 4\times 10^{-5}$. Then if $k_BT\gg \omega_o$,
Eq.(\ref{9}) collapses to $P(t) = 1-2A \sum_{k}
J_{2k+1} [ 4\tilde{\Delta}_o t] $, with $A= \pi \tilde{\Delta}_o/
(\omega_o\sqrt{2\pi N})$ (for ferritin, a $\tilde{\Delta}_o \sim 1\: MHz$ is
claimed, so $A \sim 2\times 10^{-3}$), equivalent to an absorption spectrum
(Fig.1):
\begin{equation}
\chi^{\prime \prime }(\omega ) =
{8{\tilde \Delta}_o A\over \omega \sqrt{\omega^2-16{\tilde \Delta}_o^2} }
\theta (\omega -4{\tilde \Delta}_o )\;;\;\;\;\;(\lambda \ll 1)\;,
\label{10}
\end{equation}
i.e., the spectrum of a set of biased 2-level systems with a fraction $A$
near resonance \cite{13}.
There is no dissipation, but the spectrum is not a sharp line
\cite{17}.

\vspace{-0.3cm}
\begin{figure}
\epsfxsize=3.1in
\hspace{.2em}
%%%XXX\epsffile{f1.ps}
\vspace{1.5\baselineskip}
\caption{The frequency response $\chi^{\prime \prime} (\omega )$ for the giant
spin assuming $T_2^{-1} =0$,
for the cases (a) when $\lambda = 10^{-3}$, and only
 degeneracy blocking occurs, and (b) when
$\lambda - \lambda^\prime = 10$, and topological
decoherence also plays a role.}
\end{figure}

Unfortunately in reality $T_2^{-1} \sim 10^3-10^6\:Hz$ (the strength
of the Nakamura-Suhl interaction); this causes $\epsilon$ to fluctuate in time,
with $\langle (\epsilon (t) -\epsilon (0))^2 \rangle ^{1/2} \sim
\mu \omega_oT_2^{-1}t$, over the energy range $\mu \omega_o$ of the
polarisation group in which the nuclei lie. Thus in a time
 $\tilde{\Delta}_o^{-1}$, $\epsilon (t)$ changes by $\Delta \!\epsilon \sim
\mu \omega_o (T_2 \tilde{\Delta}_o)^{-1/2}$. Coherence will be destroyed
unless $\Delta \!\epsilon \ll \tilde{\Delta}_o$, i.e., coherence requires
\begin{equation}
\tilde{\Delta}_o \gg [(\delta \omega_o )^2 T_2^{-1} ]^{-1/3} N^{1/3} \sim
T_2^{-1} N^{1/3} \;,
\label{8new}
\end{equation}
where the latter expression assumes  $T_2^{-1} \sim \delta \omega_o$, i.e.,
that $\delta \omega_o$ is mostly dipolar spreading \cite{18}.
Thus it is crucial that the spin diffusion rate $T_2^{-1}$ be kept well below
$\Delta_o$.
 In ferritin,
most nuclear spin diffusion will occur amongst the $N\sim 5000$ protons;
from (\ref{8new}) we see that coherence then requires $T_2^{-1} \ll
10^5\:Hz$, otherwise the peak in (\ref{10}) will be entirely washed out (we
are unaware of any $T_2$ measurements in ferritin).

In fact for most grains $\lambda \sim O(1)$ or greater, and the coherence
is destroyed simply by the coflipping of nuclear spins (i.e., by topological
decoherence \cite{8}). Fig.1 shows the case $\lambda =10$, appropriate
to Tb grains containing $\sim 10^3$ Tb ions.

So much for nuclear spins inside the grain - what of paramagnetic impurities
and nuclei in the surrounding substrate, which feel the dipolar field of the
grain? These dipolar fields are small compared to $\Omega_o$, but for a
mesoscopic grain, their number $N$ will be very large, and if $\lambda$ is
$\sim O(1)$ or greater, then they too will destroy coherence \cite{18b}.

(ii) \underline{Phonons \& Electrons}: In the presence of environmental spins,
electrons and phonons only further hinder coherence. Suppose
however we can {\it eliminate} nearly all such spins. Consider now the
electrons - these are particularly dangerous
 for coherence.If
grain {\it and} substrate are  conducting
a reasonably good description of the coupling to $\vec{S}$ is
\begin{equation}
H^{CC}_{eff}={1 \over 2}  \vec{S}
\cdot \hat{\vec{\sigma}}^{\alpha \beta}
\sum_{\vec{k}\vec{q}}F_{\vec{q}}c^{\dag}_{\vec{k}+\vec{q},\alpha}
c_{\vec{k},\beta}\;,
\label{11}
\end{equation}
where the form factor $F_{\vec{q}}=\int_G (d^3\vec{r}/V_o
 \:e^{i\vec{q}\cdot \vec{r}} \rho (\vec{r}) $ integrates the number density
$\rho (\vec{r})$ of grain spins over the grain volume $V_o \sim R_o^3$, and
$c^{\dag}_{\vec{k},\alpha}$ creates electron momentum and spin eigenstates;
by standard techniques \cite{15,19} we can derive a Caldeira-Leggett
spectral function $J(\omega ) = \pi \alpha_b \omega$, of Ohmic form, for
this coupling, where $\alpha_b = 2g^2S^2\langle \mid
F_{\vec{k}-\vec{k}^\prime} \mid^2 \rangle _{F.S.}$ averages $\mid
F_{\vec{k}-\vec{k}^\prime} \mid^2 $ over the Fermi surface
($\mid \vec{k}\mid $, $\mid \vec{k}^\prime \mid \sim k_F$),
and the dimensionless Kondo coupling $g=JN(0) \sim 0.1$ for metals; then
\begin{eqnarray}
\alpha_b & = & 2g^2S^2 \int_G {d^3\vec{r}d^3\vec{r}^\prime \over V_o^2}
 \left(  {\sin k_F \mid \vec{r} - \vec{r}^\prime \mid \over k_F
\mid \vec{r} - \vec{r}^\prime \mid } \right)^2 \nonumber \\
& \sim & \left( {gS \over k_FR_o }  \right)^2 \sim g^2S^{4/3} \;,
\label{12}
\end{eqnarray}
so $\alpha_b \gg 1$ for mesoscopic grains \cite{20}, completely destroying
coherence at any temperature \cite{15}. Even if the grain is insulating,
the surface spins still couple to substrate electrons, and (\ref{12}) is
replaced by a surface coupling $\alpha_s \sim g_s^2S^{2/3}$, with probably
 smaller exchange coupling $g_s$. Coherence then requires the damping
$\Gamma_s \sim 2 \pi \alpha_s k_BT$ to be considerably less
than the renormalised splitting $\Delta_r=\Delta_o
(\Delta_o/\Omega_o)^{\alpha_s/(1-\alpha_s)}$, which is hard to imagine
for any reasonable $k_BT$ unless $S\ll 100$.

The case of conducting grains on an insulating substrate is more interesting;
now the electronic states are discrete, with separation $\Delta \epsilon
\sim D /S \ge 10^4/S$ in temperature units, where $D$ is the bandwidth.
The detailed theory, involving both mesoscopic level fluctuations and surface
scattering is complicated, but the results are intuitively obvious.
If $\vec{S}$ flips, the electronic spins follow adiabatically since
$J \gg \Omega_o$, but surface spins  scatter electrons from one
internal orbital state to another. This gives an Ohmic coupling with
$\overline{\alpha}_s \sim \alpha_s/[1+\exp (\Delta \epsilon /k_BT)]$, and
a renormalized splitting $\Delta_r \sim \Delta_o$ (if $\Delta \epsilon >
\Omega_o$), or $\Delta_r \sim \Delta_o
(\Delta \epsilon /\Omega_o)^{\alpha_s}$, if $\Delta \epsilon \ll  \Omega_o$.
The damping rate is now
$\overline{\Gamma}_s \sim 2\pi \overline{\alpha}_s k_BT$, and rewriting
$\overline{\Gamma}_s$, $\Delta_r$, and $\Delta \epsilon$ in terms of
$S$, one finds almost perfect coherence provided that
\begin{equation}
{k_BT \over \Delta \epsilon } \sim S {k_BT \over D } \ll
{1 \over \ln {2\pi g^2S^{2/3} D \over S\Delta_r}}\;,
\label{13}
\end{equation}
For experiments at $mK$ temperatures, this means that electronic dissipation
effects can be ignored in such grains provided $S\le 10^5$.

For phonons the
damping rate $\Gamma_{\phi}$ is  \cite{15,19}
\begin{equation}
\Gamma_{\phi}(k_BT) \sim \left( {\Delta_o \over \Omega_o} \right)^2 J(\Delta_o)
\coth (\Delta_o/2k_BT) \;,
\label{14}
\end{equation}
where $J(\omega ) \sim C (\omega /\Theta_D)^3$; the coupling $C \sim
(KS)^2/ \Theta_D$, where $K$ is a typical anisotropy energy, and $\Theta_D$ is
the Debye temperature. For the relevant magnetoelastic interactions
the effect of phonons on tunneling from a metastable well
was investigated in \cite{21}, but here there is an
extra factor $(\Delta_o / \Omega_o)^2$ which arises \cite{19} because
 there is no
"diagonal coupling", i.e., no coupling to our $\tau_z$ from phonons
(in zero external field). Even without this factor $\Gamma_{\phi}$ is
very small (inverse minutes or hours). There will also be Ohmic
2-phonon couplings, but these give \cite{19} an even smaller damping
$\sim (k_BT/\Theta_D)^7$. Thus, as usual,  phonon effects on coherence are
negligible \cite{22}.

(iii) \underline{Real Experiments}: We now see that to see coherence one
requires, at the least (i) isotopic purification of the magnetic grains,
and  (ii)
an insulating substrate, purified of paramagnetic impurities. The grains
may be conducting if they are not too large - in any case they should
be purified of paramagnetic impurities.

These sample preparation requirements do not seem to be beyond current
techniques. This only leaves decoherence coming from the
measuring apparatus, which also interacts with the grains. One solution has
been discussed by Tesche \cite{23}, for coherence experiments
in SQUID's. Here we propose another "spin-echo"
set-up, which might be easier to
implement for magnetic grains.
If there is a sharp line in the absorption spectrum due to coherence, then
a long-lived echo signal should exist. Moreover, even if the
grains have a wide distribution of
bare tunneling rates $\Delta_o$, so $\chi^{\prime \prime }(\omega )$
is basically structureless, spin echo  and "burned hole" type experiments
may still unveil the existence of coherence \cite{24}.
For example, by
applying the standard $\pi /2$ magnetic pulse  (to generate
nondiagonal elements of the giant spin density matrix) followed by
one or more $\pi $ and $\pi /2$ pulses to reverse the phase decoherence
due to inhomogeneous broadening and/or  introduce a time delay
in the  transverse relaxation,
one can  extract the magnetic moment of the grains
and check whether it is consistent with susceptibility experiments
on the same grains. One may also study
damping mechanisms by observing the shape and amplitude
of the "burned hole" or multiple echo signal. These  depend, of course,
on the distributions of $\Delta_o$ and
$\epsilon$ and on the  decoherence mechanism -
the details will be given in a longer paper \cite{18}.

The crucial points here are (1) coherence can be found even for a spread
of $\{ \Delta_o \}$ over grains, (2) spurious peaks in
$\chi^{\prime \prime }$ can be eliminated, and (3) we have a {\it non-invasive
measurement}, where the operator
$\exp \{ -i\int H(\tau )d\tau \} $, acting between the pulses, does   all the
work. This last point is of course very important within the context of
quantum measurement theory.

This work was supported by NSERC in Canada,
by the International Science Foundation (MAA300), and by the
Russian Foundation for
Basic Research (95-02-06191a).


\begin{thebibliography}{99}
\bibitem{13}
    Notice that $\Delta N=0$ requires $N=$even integer - if $N=$odd integer
    and $\mu \ll 1$,
    degeneracy blocking completely blocks tunneling! In this paper we
    will assume even $N$; the modifications for odd $N$ are easily carried out.
\bibitem{16}
   If the $\{ \omega_k^{\perp} \}$ are important, they effectively
   increase the parameter $\lambda$  and give
    further decoherence. The $\xi_k$ play a similar role to the
    $\alpha_k$ (and $\xi_k = 2^{-1/2} \alpha_k$ for most systems), and so do
    not change the answers qualitatively; the expression including
    $\xi_k$ is too long to quote here.
\bibitem{18b}
    The most dangerous external spins are paramagnetic impurities in the
    substrate, since their dipolar coupling to $\vec{S}$ can easily
    exceed $10^8-10^9\: Hz$, giving $\alpha_k \sim 0.1$ even for a single
    impurity! Note however this coupling is much smaller for
    antiferromagnetic grains.
\bibitem{20}
    The naive estimate, that $\alpha \sim g^2S^2$, is incorrect because
     $k_FR_o \gg 1$, so that we have interference between scattering off
     different spins.
\bibitem{22}
    Phonons also broaden the discrete electron lines in a conducting
    grain by $\Gamma \sim \Theta_D (k_BT/\Theta_D)^3$, but $\Gamma \ll
\Delta \epsilon$ when $k_BT \ll \Delta \epsilon$, so this effect can also be
ignored.
\end{thebibliography}
\end{document}